# Electronic transport in locally gated graphene nanoconstrictions

Barbaros Özyilmaz, Pablo Jarillo-Herrero, Dmitri Efetov, and Philip Kim

*Department of Physics, Columbia University, New York, NY 10027, USA.*

We have developed the combination of an etching and deposition technique that enables the fabrication of locally gated graphene nanostructures of arbitrary design. Employing this method, we have fabricated graphene nanoconstrictions with local tunable transmission and characterized their electronic properties. An order of magnitude enhanced gate efficiency is achieved adopting the local gate geometry with thin dielectric gate oxide. A complete turn off of the device is demonstrated as a function of the local gate voltage. Such strong suppression of device conductance was found to be due to both quantum confinement and Coulomb blockade effects in the constricted graphene nanostructures.

Graphene[1,2,3] a recently discovered single sheet of graphite, stands out as an exceptional candidate for nanoscale electronic applications. Being a nearly perfect two dimensional electron gas, it has mobilities as high as 20000 $cm^2$/V·s, which give rise to ballistic transport on the 100 nm scale even at room temperature[4]. Furthermore, the unique "quasi-relativistic" carrier dynamics in graphene provides new transport phenomena ready to be explored for novel device applications. Many of these phenomena require lithographically patterned locally gated graphene nanostructures; examples range from Klein tunneling[5] and electron Veselago-lens[6] to spin qubits[7]. From an application point of view these new phenomena promise novel devices with strongly enhanced functionalities and novel operating principles.

Patterning graphene into nanostructures has been already demonstrated by a few groups [1,8,9,10] where interesting transport phenomena in confined graphene were observed. Other groups have also recently demonstrated the fabrication of local gate controlled graphene samples by selecting graphene flakes of random shape obtained by micromechanical extraction[11,12,13]. In this work we present a simple process which combines *both* the patterning of graphene sheets into any desired planar nanostructure *and* the local gating of the latter. Besides the abovementioned phenomena, this approach is also of interest for the fabrication of large arrays of identical graphene devices from wafer grown epitaxial graphene[14], where a global back-gate is absent and local gating offers the only way to modulate the carrier density.

Our sample fabrication process is summarized in Fig. 1. First, we deposit graphene flakes on top of an oxidized Si substrate employing mechanical exfoliation [1]. Subsequently, the location of selected flakes is determined with respect to predefined optical markers. Next, electron beam lithography (EBL) is used to pattern electric contacts to the flakes. The electron beam evaporation of Cr/Au (5/30nm) is followed by lift-off in warm acetone (Fig.

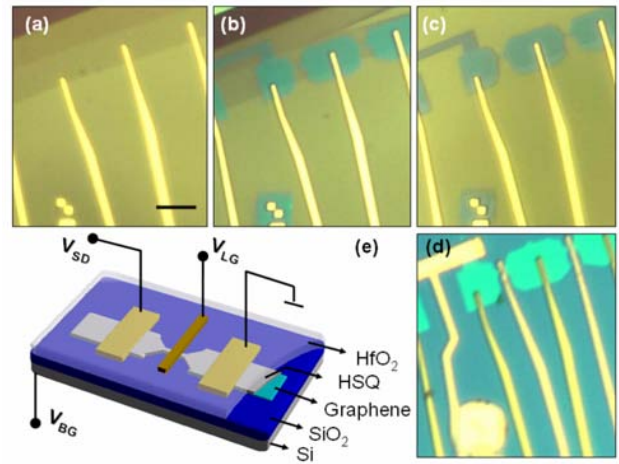

**Fig. 1**: Device fabrication and measurement scheme. (a) Optical microscope image of a graphene sample contacted by several electrodes. Scale bar 5 µm. (b) Same device as in (a) after HSQ patterning to produce a structure of bars and constrictions visible on top of graphene. (c) After etching in oxygen plasma, graphene is etched away everywhere except under the HSQ, which serves as a protective mask. (d) Picture shows several complete two-probe devices after oxygen plasma etching, ALD and local gate patterning. The metal electrodes contact the widest parts of the structure. The local gates cover the graphene constrictions and part of a graphene bar (left). (e) Schematic view of our devices. Graphene, contacted by source (S) and drain (D) electrodes, and separated from the back gate plane by 300nm $SiO_2$, and from the local gate (LG) by the top dielectric (HSQ + ALD hafnium oxide). The back gate (with voltage $V_{BG}$) is coupled to the entire graphene structure, while the local gate only couples to part of the structure.

1(a)). We then spin a thin layer (20nm) of hydrogen silsesquioxane (HSQ) solution (1:3 HSQ:MIBK)[9,10]. The latter is a high resolution negative tone electron-beam resist ideal for the reproducible patterning of graphene nanostructures down to 10nm. After resist development a short oxygen plasma step (50W, 200mTorr, 6s is enough to etch through ~10 graphene layers) is used to transfer the

HSQ pattern into the graphene sheet (Figs. 1(b)-(c)). Here HSQ acts as a protective mask, such that only the exposed graphene is etched. Without further processing steps, we deposit 15nm of high-k dielectric hafnium oxide by atomic layer deposition (ALD) directly on the samples[15]. Note that our approach does not require a nonconvalent functionalization layer[13]. Here the HSQ etch mask remains on top of the graphene device and acts simultaneously as an adhesion layer for the ALD grown dielectric. Finally, we define the local metal gates (Cr/Au (5/30nm)) using EBL (Fig. 1(d)). Thus our devices consist of a lithographically patterned graphene nanostructure sandwiched between two dielectrics, a global back gate (the highly doped Si substrate) and one or more local gates (Fig. 1(e)). Such gate configuration allows us to tune the global and local carrier densities in graphene devices via the back gate voltage ($V_{BG}$) and the local gate voltage ($V_{LG}$), respectively. The conductance, $G$, of our devices is measured at 1.7 K, as a function of $V_{BG}$ and $V_{LG}$, by using a lock-in technique with an ac excitation voltage of 100μV. The number of graphene layers in our devices is determined by Raman spectroscopy[16] and/or quantum Hall effect measurements [2,3].

Combining nanometer scaled patterning with local gate control allows us to fabricate different graphene quantum devices where the charge density varies locally. Fig. 2 shows examples of such fabricated samples ranging from graphene nanorings (lower inset in Fig. 2a) and top gated graphene Hall bars (top inset in Fig. 2(a)) to locally gated graphene nanoconstrictions (Figs. 2(a) and (b)) and ribbons (Figs. 2(a) and (c)). A typical graphene nanoconstriction is shown in Fig. 2(b). Here the width of a graphene ribbon is reduced from about 1 µm to a 30 nm wide constriction with a channel length of about 100 nm.

The conductance of bulk graphene samples remains finite at low temperatures even for zero carrier density [2,3]. This is highly undesirable for electronic devices that require an OFF state (i.e., zero conductance state), such as semiconductor transistors or quantum dots. One can, however, overcome this drawback by engineering graphene nanoconstrictions. On one hand, due to quantum confinement in the transverse direction, graphene develops a band gap in the constriction region. Alternatively, it has been suggested that small irregularities in the constriction geometry, can lead to the localization of charge in small islands, which turns into a suppression of conductance due to Coulomb blockade [17]. We note that in a continuous graphene nanostructure, the latter effect cannot occur without the formation of tunneling barriers for which a band gap due to confinement is still necessary. Therefore, in realistic samples, we expect both phenomena to take place.

A typical example of such a locally gated graphene nanoconstriction, is shown in Fig. 2(d). By tuning the local gate on top of this nanoconstriction, we can turn off the device completely ($G<10^{-10}$S), while the graphene

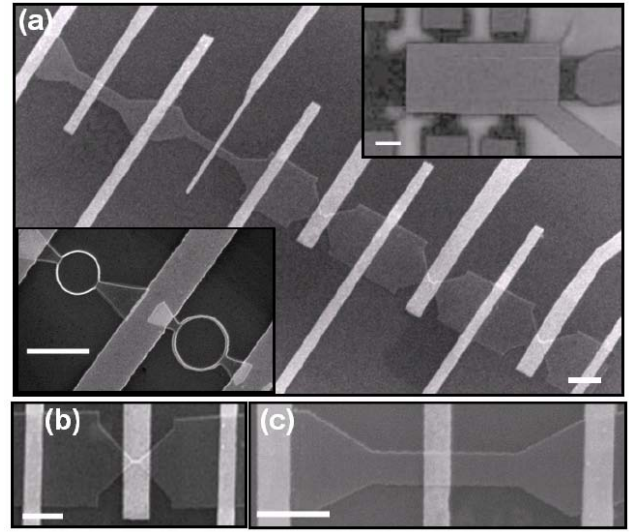

**Fig. 2:** (a) Scanning electron microscopy (SEM) picture showing several complete two-probe devices after oxygen plasma etching, ALD and local gate patterning and metal deposition. Lower left inset shows HSQ patterned rings;. Upper right inset shows a top gated bilayer graphene Hall bar. SEM picture of a locally gated graphene nanoconstriction (b) and graphene ribbon (c). All scale bars 1µm.

'electrodes' that lead to the constriction remain highly conductive ($G>e^2/h$). Figure 2(b) shows a conductance map of the same device as a function of both $V_{BG}$ and $V_{LG}$. The most notable feature is a diagonally oriented insulating region, representing the ($V_{BG}$, $V_{LG}$) range where the device is in the OFF state. Outside this region, the conductance increases rapidly, the device is turned ON. Note that, compared to previous nanoribbon devices[10], the local gate is an order of magnitude more effective in the ON-OFF modulation of the conductance. This is mainly due to the increased capacitive coupling, which is a consequence of the reduced dielectric thickness under the local gate. In addition, the fabrication of locally gated constrictions will allow realizing tunable tunnel barriers in order to study graphene quantum dots [10]. In fact, in all our nanoconstriction devices we have observed reproducible sharp peaks in the conductance as $V_{LG}$ approaches the OFF regime (see Fig. 3(a)), indicating the presence of charging effects. To gain insight into the relative contribution of quantum confinement and Coulomb blockade effects to the suppression of $G$, we have measured the stability diagram ($G$ vs ($V_{SD}$, $V_{LG}$)) for our nanoconstriction device (see Fig. 3(c)). The conductance plot shows a large central region of strongly suppressed conductance, with a series of irregular, diamond-shaped, weakly conducting regions superimposed. These irregular "Coulomb diamonds" are characteristic of multiple quantum dots in series[18], which are likely to form during the etching process (see Fig. 3(d)). A precise value of the charging energy and band gap values is difficult to obtain without a detailed knowledge of the dot configurations. However, we can roughly estimate the relative importance of the charging effects by the width of the second largest "diamond" with respect to the largest one. We estimate that the contribution of Coulomb blockade to the suppression of conductance, i.e. the ratio of charging energy to band gap due to confinement for this particular device is of order ~50 %. We note that similar features were observed in graphene nanoribbon devices controlled by a back gate only, but were analyzed only in terms of band gap formation due to confinement [10]. A more quantitative study of these two contributions will need devices where a single quantum dot is realized, for example by fabricating two smooth constrictions in series [4].

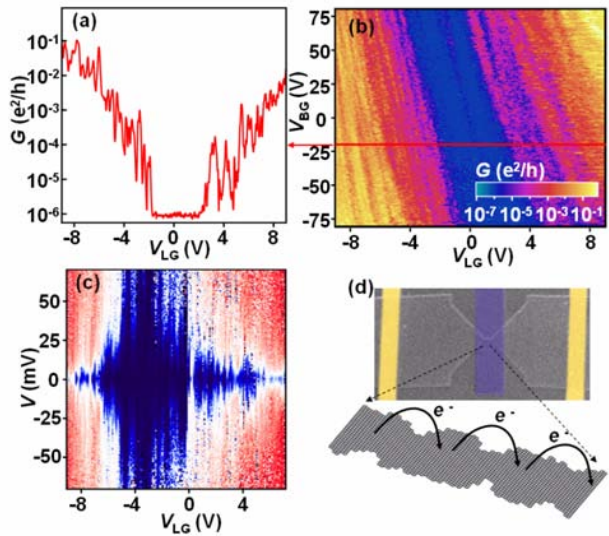

**Fig. 3:** Local gate control of electronic transport in graphene nanoconstrictions. **(a)** Conductance, $G$, in logarithmic scale, versus local gate voltage, $V_{LG}$, for one of our constrictions (horizontal trace extracted from (b)). The presence of a gap, as well as sharp peaks related to charging effects is clearly visible. (b) Color map of Log($G$) versus ($V_{LG}$, $V_{BG}$) for the constriction. The diagonal blue region corresponds to the region of zero conductance. (c) Stability diagram $G(V_{LG}, V_{SD})$ at $V_{BG} = 80V$ (dark blue is zero, red is $e^2/h$) for the same nanoconstriction. (d) False color SEM picture of the measured constriction showing S-D contacts (yellow) and the local gate (purple). The zoom-in illustrates schematically the presence of edge roughness in lithograhically pattened nanoconstrictions, which may lead to the formation of quantum dots in series. Scale bar is 1 μm.

In summary, we have demonstrated a simple approach for the fabrication and local gating of lithographically patterned graphene sheets into any planar geometry. This allows the design of graphene devices for fast exploration of novel phenomena where a local variation of the carrier density is the key to device operation. As an example we have studied graphene nanoconstrictions, where the transmission can be tuned by a local gate. The measurements reveal the importance of both quantum confinement and Coulomb blockade effects in the suppression of the conductance.

This work is supported by the ONR (N000150610138), FENA, NSF CAREER (DMR-0349232) and NSEC (CHE-0117752), and the New York State Office of Science, Technology, and Academic Research (NYSTAR).